# A counterexample to the quantum interest conjecture.

Dan Solomon
Rauland-Borg Corporation
Mount Prospect, IL
Email: dan.solomon@rauland.com
Sept. 19, 2010

**Abstract.**

According to the quantum interest conjecture any negative energy pulse must be associated with a positive energy pulse of greater magnitude than that of the negative energy pulse. In this paper we will demonstrate a counter-example to this conjecture. We will show that, for a massless scalar field in 1-1D space-time, it is possible to generate an "isolated" negative energy pulse.

**1. Introduction.**

According to the quantum interest conjecture any negative energy pulse must be associated with a positive energy pulse of greater magnitude than that of the negative energy pulse. That is, an "isolated" pulse of negative energy cannot exist. The quantum interest conjecture was originally proposed by L.H. Ford and T.A. Roman [1]. This idea has been further developed and commented on by a number of others [2-5]. The quantum interest conjecture is closely associated with the concept of quantum inequalities [6,7]. The quantum inequalities are lower bounds on the weighted average of the energy density. It has been shown that the quantum interest conjecture can be derived from the quantum inequalities [3]. Counter-examples to the quantum inequalities have been claimed in two previous papers by the author [8,9].

In this paper we will demonstrate a counter-example to the quantum interest conjecture. In the following discussion a massless scalar field in 1-1 dimensional space-time in the presence of a time varying delta function potential will be examined. The field operator $\hat{\varphi}(x,t)$ obeys the Klein-Gordon equation,

$$\frac{\partial^2 \hat{\varphi}(x,t)}{\partial t^2} - \frac{\partial^2 \hat{\varphi}(x,t)}{\partial x^2} + V(x,t)\hat{\varphi}(x,t) = 0 \qquad (1.1)$$

where $V(x,t)$ is the scalar potential. For this discussion $V(x,t)$ is given by,

$$V(x,t) = 2\lambda(t)\delta(x) \tag{1.2}$$

where $\delta(x)$ is the Dirac delta function and $\lambda(t)$ is non-negative and is specified by,

$$\lambda(t) = \begin{cases} \lambda_0 & \text{for } t < 0 \\ \lambda(t) & \text{for } 0 < t < T \\ 0 & \text{for } t > T \end{cases} \tag{1.3}$$

During the interval $0 < t < T$, $\lambda(t)$ is assumed to continuously decrease from its initial value $\lambda_0$ at $t = 0$ to its final value which is zero at $t = T$. It will be shown that during this interval a negative energy pulse is produced. The pulse is isolated and is not associated with a positive energy pulse.

**2. Mode Solutions.**

In this section we will solve Eqs. (1.1) where $V(x,t)$ is given by (1.2). The solutions are given by,

$$\hat{\varphi}(x,t) = \sum_{\omega}\left(\hat{a}_{\omega}f_{\omega}(x,t) + \hat{a}_{\omega}^{\dagger}f_{\omega}^{\dagger}(x,t)\right) \tag{2.1}$$

The $\hat{a}_{\omega}$ and $\hat{a}_{\omega}^{\dagger}$ are the annihilation and creation operators, respectively. They obey the usual relationship $\left[\hat{a}_k, \hat{a}_j^{\dagger}\right] = \delta_{jk}$ and all other commutators are equal to zero. The modes $f_{\omega}(x,t)$ satisfy,

$$\frac{\partial^2 f_{\omega}(x,t)}{\partial t^2} - \frac{\partial^2 f_{\omega}(x,t)}{\partial x^2} + 2\lambda(t)\delta(x)f_{\omega}(x,t) = 0 \tag{2.2}$$

There are both odd and even solutions to this equation. The odd solutions are of the form $e^{-i\omega t}\sin(\omega x)$. These solutions do not interest us due to the fact that they are not affected by the delta function scalar potential. This is because they are equal to zero at $x = 0$. Therefore they do not contribute to the change in the energy and will not be considered further.

Solutions to (2.2) are considered in Ref [10]. It is shown in Appendix 1 that, using the results of Ref [10], the even solutions are given by,

$$f_{\omega}(x,t) = e^{-i\omega t}\frac{A_{\omega}}{\sqrt{2\omega}}\left(\cos(\omega x) + B_{\omega}(t - |x|)e^{i\omega|x|}\right) \tag{2.3}$$

where $A_\omega$ is a normalization constant and the function $B_\omega(t)$ is given by,

$$\frac{dB_\omega(t)}{dt} + (\lambda(t) - i\omega)B_\omega(t) = -\lambda(t) \tag{2.4}$$

Initially, when $t < 0$, $\lambda(t) = \lambda_0$. For this initial static case $B_\omega(t)$ is a constant and will be written as $B_{0,\omega}$. It is evident from (2.4) that,

$$B_{0,\omega} = \frac{\lambda_0}{(i\omega - \lambda_0)} \tag{2.5}$$

Therefore the initial solution, for $t < 0$, can be written as,

$$f_\omega(x,t) = e^{-i\omega t}\chi_\omega(x) \tag{2.6}$$

where,

$$\chi_\omega(x) = \frac{A_\omega}{\sqrt{2\omega}}\left(\cos(\omega x) + B_{0,\omega}e^{i\omega|x|}\right) \tag{2.7}$$

In addition to this assume the boundary conditions $\chi_\omega(\pm L/2) = 0$ where $L \to \infty$. From these boundary conditions we obtain $\cos(\omega L/2) + B_{0,\omega}e^{i\omega L/2} = 0$. This yields,

$$\omega\cos(\omega L/2) = -\lambda_0 \sin(\omega L/2) \tag{2.8}$$

The solutions to this equation are given by,

$$\omega_n = \omega_{0,n} + \frac{2\delta_n}{L} \text{ where } \omega_{0,n} = \frac{2\pi}{L}\left(n + \frac{1}{2}\right) \text{ with } n = 0,1,2,\ldots \tag{2.9}$$

Use this in (2.8) along with some trig identities and the fact that $\cos(\omega_0 L/2) = 0$ to obtain,

$$\omega_n \sin\delta_n = \lambda_0 \cos\delta_n \tag{2.10}$$

The normalization constant $A_\omega$ is given by solving the normalization condition,

$$1 = -i\int_{-(L/2)}^{+(L/2)}\left(f_\omega \frac{\partial f_\omega^*}{\partial t} - f_\omega^* \frac{\partial f_\omega}{\partial t}\right)dx = A_\omega^2 \int_{-(L/2)}^{+(L/2)}\left|\cos(\omega x) + B_0 e^{i\omega|x|}\right|^2 dx \tag{2.11}$$

From this we obtain,

$$A_\omega^2 = \frac{2}{L}\left(1 + \frac{2\lambda_0}{(\lambda_0^2 + \omega^2)L}\right)^{-1} \tag{2.12}$$

## 3. Kinetic energy density.

The kinetic energy density operator is given by,

$$\hat{T}_{00}(x,t) = \frac{1}{2}\left(\left|\frac{\partial \hat{\varphi}(x,t)}{\partial t}\right|^2 + \left|\frac{\partial \hat{\varphi}(x,t)}{\partial x}\right|^2\right) \tag{3.1}$$

The kinetic energy density operator is the same as the energy density operator at any point in space where the scalar potential is zero. This is due to the fact that the energy density operator includes a term containing the scalar potential. For the problem we are considering the scalar potential is zero everywhere except at $x=0$. This means that the kinetic energy density and energy density operators are identical except at this point. When $t>T$ the kinetic energy density and energy density are identical everywhere because, from (1.3), the scalar potential is zero.

In order to calculate the kinetic energy density expectation value, the state vector on which the field operators act must be specified. This state vector will be designated by $|0\rangle$ and is defined by the relationship $\hat{a}_\omega|0\rangle = 0$.

From the above discussion the kinetic energy density expectation value is,

$$T_{00}(x,t) = \langle 0|\hat{T}_{00}(x,t)|0\rangle \tag{3.2}$$

This can be shown to be equal to,

$$T_{00}(x,t) = \sum_{n=0}^{\infty} \varepsilon_{\omega_n}(x,t) \tag{3.3}$$

where,

$$\varepsilon_\omega(x,t) = \frac{1}{2}\left(\left|\frac{\partial f_\omega(x,t)}{\partial t}\right|^2 + \left|\frac{\partial f_\omega(x,t)}{\partial x}\right|^2\right) \tag{3.4}$$

and where $\varepsilon_\omega$ is the kinetic energy density of a given mode.

There is a problem with evaluating (3.3) which is due to the fact it will be infinite. However we are not really interested in the total energy density but only the change in the energy density with respect to the unperturbed vacuum state. To achieve this result we will use mode regularization. The kinetic energy density of each mode will be given by,

$$\varepsilon_{\omega,R}(x,t) = \varepsilon_\omega(x,t) - \varepsilon_{0,\omega_0} \tag{3.5}$$

with $\varepsilon_{0,\omega_0}$ being the kinetic energy density of the unperturbed state and is given by $\varepsilon_{0,\omega_0} = \omega_0/2L$. The total regularized kinetic energy density is given by summing up the change in the kinetic energy density of each mode to obtain,

$$T_{00,R}(x,t) = \sum_{n=0}^{\infty} \varepsilon_{\omega_n,R}(x,t) \tag{3.6}$$

Use (2.6) and (2.7) in (3.4) to obtain,

$$\varepsilon_\omega(x,t) = \frac{A_\omega^2 \omega}{4}\left(1+\left(B_{0,\omega} + B_{0,\omega}^*\right) + 2\left|B_{0,\omega}\right|^2\right) \tag{3.7}$$

This equation is derived in the Appendix 3. Note that $\varepsilon_\omega(x,t)$ is both space and time independent. Therefore we will drop the explicit dependence on $x$ and $t$ in the rest of this section.

Next, use (2.5) in the above to obtain,

$$\varepsilon_\omega = \frac{A_\omega^2 \omega}{4} \tag{3.8}$$

The "regularized" kinetic energy of each mode is, then, given by,

$$\varepsilon_{R,\omega_n} = \frac{A_{\omega_n}^2 \omega_n}{4} - \frac{\omega_{0,n}}{2L} \tag{3.9}$$

Use this in (3.6) to obtain,

$$T_{00,R} = \sum_{n=0}^{\infty} \varepsilon_{R,\omega_n} \tag{3.10}$$

In the limit that $L \to \infty$ we can use $\sum_{n=0}^{\infty} \to \int_0^\infty \frac{L d\omega_0}{2\pi}$ to obtain,

$$T_{00,R} = \frac{L}{2\pi}\int_0^\infty \varepsilon_{R,\omega_n} d\omega_0 \tag{3.11}$$

This is evaluated in Appendix 2 to obtain,

$$T_{00,R} = \frac{\lambda_0}{2\pi L} \tag{3.12}$$

The total kinetic energy, $E_K$, is the kinetic energy density, $T_{00,R}$, integrated over all space to obtain,

$$E_K = \int_{-(L/2)}^{+(L/2)} T_{00,R} dx = \frac{\lambda_0}{2\pi} \qquad (3.13)$$

From the above results we see that the kinetic energy density, $T_{00,R}$, is effectively zero everywhere in the limit $L \to \infty$. However the total kinetic energy, $E_K$, is positive.

**4. Generating a negative energy pulse.**

At this point we have established the initial state which is valid for $t < 0$. Between $t = 0$ and $t = T$ the scalar potential decreases to zero. In the following discussion we will show that this reduction in the scalar potential results in the launching of a pulse of negative kinetic energy density. To see why this makes sense consider the expression for the total kinetic energy $\xi_\omega(t)$ associated with a given mode. This is just the kinetic energy density of the mode integrated over all space. Referring to (3.4) and (3.5) we obtain,

$$\xi_\omega(t) = \int \varepsilon_{\omega,R}(x,t)dx = \int \left\{ \frac{1}{2}\left( \left|\frac{\partial f_\omega(x,t)}{\partial t}\right|^2 + \left|\frac{\partial f_\omega(x,t)}{\partial x}\right|^2 \right) - \varepsilon_{0,\omega_0} \right\} dx \qquad (4.1)$$

where integration from $-L/2$ to $+L/2$ is implied. Take the time derivative of this expression and use (2.2) to obtain,

$$\frac{\partial \xi_\omega(t)}{\partial t} = \frac{1}{2}\int \left\{ \frac{\partial}{\partial x}\left( \frac{\partial f_\omega(x,t)}{\partial x}\frac{\partial f_\omega^*(x,t)}{\partial t} + \frac{\partial f_\omega(x,t)}{\partial t}\frac{\partial f_\omega^*(x,t)}{\partial x} \right) - 2\lambda(t)\delta(x)\frac{\partial |f_\omega(x,t)|^2}{\partial t} \right\} dx$$

(4.2)

This yields,

$$\frac{\partial \xi_\omega(t)}{\partial t} = -\lambda(t)\frac{\partial |f_\omega(0,t)|^2}{\partial t} \qquad (4.3)$$

where we have used the boundary condition $\chi_\omega(\pm L/2) = 0$. Next consider what happens when $\lambda(t)$ decreases from $\lambda_0$ at $t = 0$ to zero at $t = T$. The total change in the kinetic energy of the mode during this time interval is given by integrating the above quantity to obtain,

$$\Delta \xi_\omega (0 \to T) = -\int_0^T \lambda(t) \frac{\partial |f_\omega(0,t)|^2}{\partial t} dt \tag{4.4}$$

What might this energy be? Assume that $\lambda(t)$ decreases quasi-statically. That is, we reduce $\lambda(t)$ slowly enough so that $f_\omega(0,t)$ is approximated by the static solution given by equation (2.6) and (2.7). Technically this assumption must be verified but for the moment let us assume that it is "reasonable". Use (2.6) along with (2.7) and (2.5) and replace $\lambda_0$ by $\lambda(t)$ to obtain,

$$|f_\omega(0,t)|^2 \cong \frac{A_\omega^2}{2\omega}|1+B_{0,\omega}|^2 = \frac{\omega A_\omega^2}{2(\omega^2 + \lambda(t)^2)} \tag{4.5}$$

Therefore,

$$\frac{\partial |f_\omega(0,t)|^2}{\partial t} \cong -\frac{\lambda(t)\omega A_\omega^2}{(\omega^2 + \lambda(t)^2)^2} \frac{d\lambda(t)}{dt} \tag{4.6}$$

Use this in (4.4) to obtain,

$$\Delta \xi_\omega (0 \to T) \cong \int_0^T \frac{\lambda(t)^2 \omega A_\omega^2}{(\omega^2 + \lambda(t)^2)^2} \frac{d\lambda(t)}{dt} dt \tag{4.7}$$

Due to the fact that $\lambda(t)$ is decreasing the quantity $d\lambda(t)/dt$ will be negative which means that $\Delta\xi_\omega(0 \to T)$ will also be negative. The change in the total kinetic energy is summation of the change in the kinetic energy of each mode. This yields,

$$\Delta E_K (0 \to T) = \sum_n \Delta \xi_{\omega_n} (0 \to T) \tag{4.8}$$

This is negative because each $\Delta \xi_{\omega_n}(0 \to T)$ is negative.

What happens to the energy density during this time interval? Note that at $t=T$ the energy density in the region outside of $|x|<T$ cannot have changed because the effect due to the changing of the scalar potential cannot travel faster than the speed of light which we have taken to be equal to 1. Therefore all the energy generated during the interval $T > t > 0$ is contained within the region $|x| < T$ at the time $t = T$. This means that the average energy density within the region $|x| < T$ is negative.

Next what happens for the time $t > T$ for which $\lambda(t) = 0$? Referring to (4.3) the total kinetic energy density can no longer change. Therefore the negative energy that was produced during the time interval $T > t > 0$ will move out at the speed of light and not be followed by a pulse of positive energy since, for $t > T$, no further energy (negative or positive) is being generated. The result is an isolated region of negative energy moving away from the origin at the speed of light. One half of the negative energy moves in the positive direction and other half moves in the negative direction.

These results are based on the validity of the quasi-static approximation. They will be verified in the next two sections when we consider exact solutions.

## 5. An exact solution I.

In this section we will show that a negative energy pulse is radiated for the case where $\lambda(t)$ is given by,

$$\lambda(t) = \begin{cases} \lambda_0 \text{ for } t < 0 \\ \lambda_0/2 \text{ for } 0 \leq t < T \\ 0 \text{ for } t > T \end{cases} \quad (5.1)$$

Define,

$$C_\omega(t) = 1 + B_\omega(t) \quad (5.2)$$

Use this in (2.3) to obtain,

$$f_\omega(0,t) = e^{-i\omega t} \frac{A_\omega}{\sqrt{2\omega}} C_\omega(t) \quad (5.3)$$

Use this result and (5.1) in (4.4) to show that the change in kinetic energy during the interval $t = 0$ to $t = T$ for a given mode is given by,

$$\Delta \xi_\omega(0 \to T) = -\frac{\lambda_0}{2}\left(\frac{A_\omega}{\sqrt{2\omega}}\right)^2 \int_0^T \frac{\partial |C_\omega(t)|^2}{\partial t} dt = -\frac{\lambda_0}{2}\left(\frac{A_\omega}{\sqrt{2\omega}}\right)^2 \left(|C_\omega(T)|^2 - |C_\omega(0)|^2\right) \quad (5.4)$$

From (2.4) we obtain,

$$\frac{dC_\omega(t)}{dt} + (\lambda(t) - i\omega) C_\omega(t) = -i\omega \quad (5.5)$$

For $t < 0$ the mode solutions are given by (2.6). For this case $B_\omega(t) = B_{0,\omega}$ where $B_{0,\omega}$ is given by (2.5). Therefore for $t < 0$,

$$C_\omega(t) \underset{t<0}{=} C_{0,\omega} = 1 + B_{0,\omega} = i\omega/(i\omega - \lambda_0) \tag{5.6}$$

For $t > 0$ the solution to (5.5) is,

$$C_\omega(t) = e^{i\omega t} e^{-F(t)} \left( C_\omega(0) - i\omega \int_0^t e^{-i\omega t'} e^{F(t')} dt' \right) \tag{5.7}$$

where $C_\omega(0) = C_{0,\omega}$ and,

$$F(t) = \int_0^t \lambda(u) du \tag{5.8}$$

For the interval $T > t > 0$ we use $\lambda(t) = \lambda_0/2$ in (5.7) to obtain,

$$C_\omega(t) \underset{T>t>0}{=} \left\{ \left( C_{0,\omega} + \frac{i\omega}{(\lambda_0/2) - i\omega} \right) e^{i\omega t} e^{-\lambda_0 t/2} + \frac{i\omega}{i\omega - (\lambda_0/2)} \right\} \tag{5.9}$$

Consider the case where $T = 100/\lambda_0$. In this case $e^{-\lambda_0 T/2} = e^{-50}$. Use this in (5.9) to obtain,

$$C_\omega(T) \cong \frac{i\omega}{i\omega - (\lambda_0/2)} \tag{5.10}$$

Use this in (5.4) to obtain,

$$\Delta \xi_\omega(0 \to T) \cong -\frac{\lambda_0 A_\omega^2}{4\omega} \left( \left| \frac{i\omega}{i\omega - (\lambda_0/2)} \right|^2 - \left| \frac{i\omega}{i\omega - \lambda_0} \right|^2 \right) \tag{5.11}$$

Rearrange terms to obtain,

$$\Delta \xi_\omega(0 \to T) \cong -\frac{\lambda_0 A_\omega^2 \omega^2}{4\omega} \left( \frac{3(\lambda_0/2)^2}{\left(\omega^2 + (\lambda_0/2)^2\right)\left(\omega^2 + \lambda_0^2\right)} \right) \tag{5.12}$$

Therefore during the interval from $t = 0$ to $t = T$ the change in the kinetic energy of each mode is negative. For $t > T$ when $\lambda(t) = 0$ there is no additional change in the kinetic energy. The result is a pulse of negative energy moving out at the speed of light. This can be seen by direct examination of the kinetic energy density.

Using (2.3) and (2.4) in (3.4) the kinetic energy density of each mode, for $t > 0$ and $x > 0$ is,

$$\varepsilon_\omega(x,t) = \frac{A_\omega^2}{4\omega}\left(\omega^2 - i\omega\lambda(x-t)\left(C_\omega(x-t) - C_\omega^*(x-t)\right) + 2\lambda^2(x-t)|C_\omega(x-t)|^2\right) \quad (5.13)$$

The details of this calculation are given in Appendix 3. The regularized energy density of each mode is,

$$\varepsilon_{R,\omega}(x,t) = \varepsilon_\omega(x,t) - \varepsilon_{0,\omega_0} \quad (5.14)$$

Use (5.13) to obtain,

$$\varepsilon_{R,\omega}(x,t) = \delta\varepsilon_\omega + \Delta\varepsilon_\omega(t-x) \quad (5.15)$$

where,

$$\Delta\varepsilon_\omega(t-x) = \frac{A_\omega^2}{4\omega}\left(\begin{array}{l} -i\omega\lambda(t-x)\left(C_\omega(t-x) - C_\omega^*(t-x)\right) \\ +2\lambda^2(t-x)|C_\omega(t-x)|^2 \end{array}\right) \quad (5.16)$$

and,

$$\delta\varepsilon_\omega = \frac{A_\omega^2 \omega_n}{4} - \frac{\omega_0}{2L} \quad (5.17)$$

$\delta\varepsilon_\omega$ is identical to the kinetic energy density of a mode in the static case (see Eq. (3.9)). Since, from the discussion in Appendix 2 $\delta\varepsilon_\omega \to O(1/L^2)$ (see Eq. (8.5)) it can be removed from the equation in the limit $L \to \infty$. Therefore we obtain,

$$\varepsilon_{R,\omega}(x,t) = \Delta\varepsilon_\omega(t-x) \quad (5.18)$$

As can be seen by this expression each mode consists of energy radiating out at the speed of light. Recall that expression applies to the positive half-plane. There is an analogous expression for the negative half-plane.

Using this result the total kinetic energy density is,

$$T_{00,R}(x,t) = \frac{L}{2\pi}\int_0^\infty \Delta\varepsilon_\omega(t-x)d\omega_0 \quad (5.19)$$

We can drop all $O(1/L^2)$ terms from the integrand in the above expression. This allows us to write $\Delta\varepsilon_\omega(x,t)$ as $\Delta\varepsilon_{\omega_0}(x,t)$. That is, we can replace $\omega$ by $\omega_0$. Thus we obtain,

$$T_{00,R}(x,t) = \frac{L}{2\pi}\int_0^\infty \Delta\varepsilon_\omega(t-x)d\omega \qquad (5.20)$$

where we have replaced the dummy variable $\omega_0$ with $\omega$ to simplify notation. Note that for this case, in the limit $L \to \infty$, $A_\omega^2 \to 2/L$ in (5.16).

## 6. An exact solution II.

In this section we will solve for the energy density for the case where the scalar potential decreases continuously to zero. Let $\lambda(t)$ be given by,

$$\lambda(t) = \begin{cases} \lambda_0 & \text{for } t < 0 \\ \dfrac{f_2}{1+f_2 t} - (f_2 - \lambda_0) & \text{for } 0 \le t < T \\ 0 & \text{for } t > T \end{cases} \qquad (6.1)$$

where,

$$T = \frac{\lambda_0}{f_2(f_2 - \lambda_0)} \qquad (6.2)$$

Note that $\lambda(t)$ decreases continuously from $\lambda_0$ at $t=0$ to zero at $t=T$. Use (6.1) in (5.8) to obtain,

$$F(t) = \ln(1+f_2 t) - f_3 t \qquad (6.3)$$

where $f_3 = f_2 - \lambda_0$. Use this result in (5.7) to obtain,

$$C_\omega(t) = \frac{e^{i\omega t} e^{f_3 t}}{(1+tf_2)}(C_\omega(0) - i\omega G_\omega(t)) \text{ for } 0 < t < T \qquad (6.4)$$

where, $C_\omega(0) = 1 + B_{0,\omega}(0) = i\omega/(i\omega - \lambda_0)$ and,

$$G_\omega(t) = \frac{1-e^{-(i\omega+f_3)t}}{(i\omega+f_3)}\left[1 + \frac{f_2}{(i\omega+f_3)}\right] - f_2\left(\frac{te^{-(i\omega+f_3)t}}{(i\omega+f_3)}\right) \qquad (6.5)$$

As discussed in the previous section energy will be radiated during the time that $\lambda(t)$ is reduced to zero during the interval $0 \le t < T$. At time $t = T$ this energy will be located in the region for which $|x| < T$.

What is the total kinetic energy at time $t = T$ between points $x = 0$ and $x = T$? In the previous section we have used (4.3) to determine the energy per mode that is generated when the potential is decreased. Here we will calculate the energy directly be integrating the energy density at time $t = T$, $T_{00,R}(x,T)$, between $x = 0$ and $x = T$. We obtain,

$$E(0 \to T) = \int_0^T T_{00,R}(x,T) dx \tag{6.6}$$

Use (5.20) in the above to obtain,

$$E(0 \to T) = \int_0^T dx \frac{L}{2\pi} \int_0^\infty \Delta \varepsilon_\omega (T-x) d\omega \tag{6.7}$$

Making the substitution $u = T - x$ we obtain,

$$E(0 \to T) = \int_0^T du \int_0^\infty \Delta \bar{\varepsilon}_\omega (u) d\omega \tag{6.8}$$

where $\Delta \bar{\varepsilon}_\omega (u) = \frac{L}{2\pi} \Delta \varepsilon_\omega (u)$. Refer to (5.16) and substitute $A_\omega^2 = 2/L$ in the limit $L \to \infty$ we obtain,

$$\Delta \bar{\varepsilon}_\omega (u) = \frac{1}{4\pi\omega} \left( -i\omega \lambda(u) \left( C_\omega(u) - C_\omega^*(u) \right) + 2\lambda^2(u) |C_\omega(u)|^2 \right) \tag{6.9}$$

We use this result in (6.8) which is evaluated using numerical integration as discussed in Appendix 4. Some results are shown in the following table.

| $f_2$ | $\lambda_0$ | $E(0 \to T)$ |
|---|---|---|
| 1 | 1/2 | −0.0134 |
| 2 | 1/2 | −0.00413 |
| 2 | 1 | −0.0268 |
| 4 | 2 | −0.0536 |
| 4 | 3 | −0.155 |

It can be seen that $E(0 \to T)$ is negative which is consistent with the analysis of Sections 4 and 5. $T$ can be found by using the above values in (6.2).

The above result has been obtained by setting $x > 0$ in (5.13). However the same result would is obtained for $x < 0$. Therefore at this point we have found that at $t = T$ the total energy in the region $T > |x|$ is negative. The average energy density in the region is $E(0 \to T)/T$ and is negative. In the region $|x| > T$ the energy density is effectively zero.

When $t > T$ the quantity $\lambda(t) = 0$ as specified in Eq. (6.1). The result of this is that no additional energy is produced. The kinetic energy density for the region $t - x > T$ is zero for the half space where $x > 0$. This can be seen by referring to (5.16) which shows that the kinetic energy density of a given mode is zero when $\lambda(t - x) = 0$ is zero which will be the case when $t - x > T$. There is an analogous result for the half space $x < 0$.

At this point we will summarize the above results. For $t < 0$ the kinetic energy density is effectively zero over all space. By "effectively" we mean it goes as $1/L$ as $L \to \infty$. From $t = 0$ to $t = T$ the scalar potential is decreasing to zero and a negative energy pulse is being radiated and propagating outward at the speed of light. For $t > T$ there is no further change in the total kinetic energy because $\lambda(t) = 0$ for $t > T$. The final result for $t = t_f$ where $t_f > T$ is that the kinetic energy density is effectively zero over the region $|x| > t_f$ (because in this region $t_f - |x| < 0$ and the radiated pulse hasn't reached here yet), it has an average negative value over the region $t_f > |x| > t_f - T$ (because in this region $0 < t_f - |x| < T$ and $\lambda(t_f - |x|)$ is decreasing), and it is zero in the region $t_f - T > |x|$ (because in this region $t_f - |x| > T$ and, therefore, $\lambda(t_f - |x|)$ will be zero). This means that an isolated pulse of negative energy is moving in the positive x-direction at the speed of light and an equivalent negative energy pulse is moving in the negative x-direction.

In conclusion we have shown that it is possible to violate the quantum interest conjecture and generate a pulse of negative energy that is not associated with a positive energy pulse.

**Appendix 1.**

In the following discussion we refer to Eqs. 5.4 and 5.5 of Ref. [10]. We have made some modifications to be consistent with the notation of the present paper.

According to Eq. 5.4 of Ref. [10] for the zero mass scalar field considered here $f_\omega(x,t)$ is given by,

$$f_\omega(x,t) = N_\omega e^{-i\omega t} \cos(\omega x) - \int_{-\infty}^{t-|x|} \lambda(\tau) f_\omega(0,\tau) d\tau \tag{7.1}$$

where $N_\omega = A_\omega/\sqrt{2\omega}$. (Note – there is a minor error in Ref [10] which is corrected here. In [10] the upper limit is the integral is $t-x$. The correct value is $t-|x|$). Set $x=0$ in the above and take the derivative with respect to time of both sides of the equation to obtain,

$$\frac{d}{dt}\left(f_\omega(0,t) - N_\omega e^{-i\omega t}\right) = -\lambda(t) f_\omega(0,t) \tag{7.2}$$

Use this in (7.1) to obtain,

$$f_\omega(x,t) = N_\omega e^{-i\omega t} \cos(\omega x) + \left(f_\omega(0,t-|x|) - N_\omega e^{-i\omega(t-|x|)}\right) \tag{7.3}$$

where we have used $\left(f_\omega(0,-\infty) - N_\omega e^{-i\omega(-\infty)}\right) = 0$. Define,

$$f_\omega(0,t) = N_\omega C_\omega(t) e^{-i\omega t} \tag{7.4}$$

and use this in (7.2) to obtain,

$$\frac{\partial C_\omega(t)}{\partial t} + \left(\lambda(t) - i\omega\right) C_\omega(t) = -i\omega \tag{7.5}$$

Use (7.4) in (7.3) to obtain,

$$f_\omega(x,t) = N_\omega e^{-i\omega t} \left[\cos(\omega x) + \left(C_\omega(t-|x|) - 1\right) e^{i\omega|x|}\right] \tag{7.6}$$

Define $B_\omega(t) = C_\omega(t) - 1$ and use this in (7.5) and (7.6) to obtain (2.3) and (2.4).

The same result is also obtained in Ref. [11] using a different approach.

**Appendix 2.**

To evaluate (3.11) we will first expand $\varepsilon_{R,\omega_n}$ in terms of $1/L$. To do this refer to (2.12) to obtain,

$$A_\omega^2 = \frac{2}{L} - \frac{4\lambda}{(\lambda^2 + \omega^2)L^2} + O(1/L^3) \tag{8.1}$$

Use this in (3.9) to obtain,

$$\varepsilon_{R,\omega_n} = \frac{1}{2L}(\omega_n - \omega_{0,n}) - \frac{\omega_n}{L^2}\left(\frac{\lambda_0}{\lambda_0^2 + \omega_n^2}\right) + O(1/L^3) \tag{8.2}$$

Use (2.10) to obtain,

$$\delta_n = \arcsin\left(\frac{\lambda_0}{\sqrt{\lambda_0^2 + \omega_n^2}}\right) = \arcsin\left(\frac{\lambda_0}{\sqrt{\lambda_0^2 + \omega_{0,n}^2}}\right) + O(1/L) \tag{8.3}$$

From this and (2.9) we obtain,

$$\omega_n = \omega_{0,n} + \frac{2}{L}\arcsin\left(\frac{\lambda_0}{\sqrt{\lambda_0^2 + \omega_{0,n}^2}}\right) + O(1/L^2) \tag{8.4}$$

Use this in (8.2) to yield,

$$\varepsilon_{R,\omega_n} = \frac{1}{L^2}\arcsin\left(\frac{\lambda_0}{\sqrt{\lambda_0^2 + \omega_{0,n}^2}}\right) - \frac{\omega_{0,n}}{L^2}\left(\frac{\lambda_0}{\lambda_0^2 + \omega_{0,n}^2}\right) + O(1/L^3) \tag{8.5}$$

Use this in (3.11) to obtain,

$$T_{00,R} = \frac{1}{2\pi L}\int_0^\infty \left(\arcsin\left(\frac{\lambda_0}{\sqrt{\lambda_0^2 + \omega_0^2}}\right) - \left(\frac{\lambda_0 \omega_0}{\lambda_0^2 + \omega_0^2}\right)\right) d\omega_0 + O(1/L^2) \tag{8.6}$$

To evaluate this first note that the integrand goes to zero sufficiently fast as $\omega_0 \to \infty$ to obtain,

$$T_{00,R} \underset{\Lambda \to \infty}{=} \frac{1}{2\pi L}\int_0^\Lambda \left(\arcsin\left(\frac{\lambda_0}{\sqrt{\lambda_0^2 + \omega_0^2}}\right) - \left(\frac{\lambda_0 \omega_0}{\lambda_0^2 + \omega_0^2}\right)\right) d\omega_0 \tag{8.7}$$

where the $O(1/L^2)$ term has been dropped. The first term in the integrand can be evaluated using integration by parts to obtain,

$$\int_0^\Lambda \arcsin\left(\frac{\lambda_0}{\sqrt{\lambda_0^2+\omega_0^2}}\right)d\omega_0 = \Lambda\arcsin\left(\frac{\lambda_0}{\sqrt{\lambda_0^2+\Lambda^2}}\right)+\int_0^\Lambda \left(\frac{\lambda_0\omega_0}{\lambda_0^2+\omega_0^2}\right)d\omega_0 \qquad (8.8)$$

Use this in (8.7) to obtain,

$$T_{00,R} \underset{\Lambda\to\infty}{=} \frac{\Lambda}{2\pi L}\arcsin\left(\frac{\lambda_0}{\sqrt{\lambda_0^2+\Lambda^2}}\right) = \frac{\lambda_0}{2\pi L} \qquad (8.9)$$

**Appendix 3.**

Calculate the kinetic energy density for a given mode in the region $x>0$. The energy density $\varepsilon_\omega(x,t)$ is given by (3.4). In order to determine this in the region $x>0$ refer to (2.3) to obtain,

$$\frac{\partial f_\omega}{\partial t} = \frac{A_\omega e^{-i\omega t}}{\sqrt{2\omega}}\left[-i\omega\cos(\omega x)+\left(-i\omega B_\omega(t-x)+\frac{\partial B_\omega(t-x)}{\partial t}\right)e^{i\omega x}\right] \qquad (9.1)$$

and,

$$\frac{\partial f_\omega}{\partial x} = \frac{A_\omega e^{-i\omega t}}{\sqrt{2\omega}}\left[-\omega\sin(\omega x)-\left(\frac{\partial B_\omega(t-x)}{\partial t}-i\omega B_\omega(t-x)\right)e^{i\omega x}\right] \qquad (9.2)$$

where we have used $\partial B_\omega(t-x)/\partial x = -\partial B_\omega(t-x)/\partial t$. Use (2.4) to obtain,

$$\frac{dB_\omega(t-x)}{dt}-i\omega B_\omega(t-x) = -\lambda(t-x)(1+B_\omega(t-x)) \qquad (9.3)$$

Use this in (9.1) and (9.2) to obtain,

$$\frac{\partial f_\omega}{\partial t} = \frac{A_\omega e^{-i\omega t}}{\sqrt{2\omega}}\left[-i\omega\cos(\omega x)-\lambda(t-x)(1+B_\omega(t-x))e^{i\omega x}\right] \qquad (9.4)$$

and,

$$\frac{\partial f_\omega}{\partial x} = \frac{A_\omega e^{-i\omega t}}{\sqrt{2\omega}}\left[-\omega\sin(\omega x)+\lambda(t-x)(1+B_\omega(t-x))e^{i\omega x}\right] \qquad (9.5)$$

Use these in (3.4) to obtain,

$$\varepsilon_\omega(x,t) = \frac{A_\omega^2}{4\omega}\left(\omega^2 - i\omega\lambda(x-t)(B_\omega(x-t)-B_\omega^*(x-t))+2\lambda^2(x-t)|1+B_\omega(x-t)|^2\right) \qquad (9.6)$$

Define $C_\omega(t) = 1+B_\omega(t)$ in the above to obtain Eq. (5.13).

For the static case, where $dB_\omega/dt = 0$, refer to (9.3) to obtain, $i\omega B_{0,\omega} = \lambda_0(1+B_{0,\omega})$. Use this and make the substitutions $\lambda(x-t) \to \lambda_0$ and $B_\omega(x-t) \to B_{0,\omega}$ to obtain (3.7).

**Appendix 4.**

Below is the Mathematica source code that is used to evaluate Eq. (6.8). Here the dummy variable of integration is 't' instead of 'u'. The quantity eE stands $E(0 \to T)$. The quantity energy[w,t] stands for $\pi \Delta \bar{\varepsilon}_\omega(u)$ where $\Delta \bar{\varepsilon}_\omega(u)$ is defined in (6.9). Also cC[w,t] stands for $C_\omega(t)$ which is specified in (6.4). In addition cC0[w] stands for $C_\omega(0)$ and gG[w,t] stands for $G_\omega(t)$ which is given by (6.5). The upper limit of the integral has been truncated to 100 (from infinity) because the integrand converges sufficiently fast.

```
(* Counter-example to the quantum interest conjecture*)
(* Sept 8, 2010*)
Clear[f2];Clear[lamb0];Clear[w];Clear[t];
f2=2; lamb0=1;
lambda[t]:= (f2/(1+t*f2))-(f2-lamb0);
tT=lamb0/(f2*(f2-lamb0));
f3=( f2-lamb0);
aa[w]:=I*w + f3;
exp[w,t]:= Exp[-aa[w]*t];
gG1[w,t]:= ((1-exp[w,t])/aa[w])*(1+(f2/aa[w]));
gG2[w,t]:= f2*(t*exp[w,t])/aa[w];
gG[w,t]:= gG1[w,t]-gG2[w,t];
cC0[w]:=I*w/(I*w-lamb0);
cC[w,t]:=(Exp[aa[w]*t]/(1+t*f2))*(cC0[w]-I*w*gG[w,t]);
conjC[w,t]:=Conjugate[cC[w,t]];
energy1[w,t]:=-lambda[t]*I*w*(cC[w,t]-conjC[w,t]);
energy2[w,t]:=2*(lambda[t]^2)*cC[w,t]*conjC[w,t];
energy[w,t]:=(1/(4*w))*(energy1[w,t]+energy2[w,t]);
eE =
(1/Pi)*NIntegrate[energy[w,t],{t,0,tT},{w,0,100},WorkingPre
cision→15]
```

**References.**